# Polarizing a stored proton beam by spin flip?

D. Oellers[1,2], L. Barion[3], S. Barsov[4], U. Bechstedt[1,2], P. Benati[3], S. Bertelli[3],
D. Chiladze[5], G. Ciullo[3], M. Contalbrigo[6], P.F. Dalpiaz[3], J. Dietrich[1,2],
N. Dolfus[1,2], S. Dymov[7,8], R. Engels[1,2], W. Erven[9,2], A. Garishvili[5,7],
R. Gebel[1,2], P. Goslawski[10], K. Grigoryev[4,1,2], H. Hadamek[1,2],
A. Kacharava[1,2], A. Khoukaz[10], A. Kulikov[8], G. Langenberg[1,2],
A. Lehrach[1,2], P. Lenisa[3], N. Lomidze[5], B. Lorentz[1,2], G. Macharashvili[8],
R. Maier[1,2], S. Martin[1,2], S. Merzliakov[8,1,2], I.N. Meshkov[8], H.O. Meyer[11],
M. Mielke[10], M. Mikirtychiants[4,1,2], S. Mikirtychiants[4,1,2], A. Nass[7],
M. Nekipelov[1,2], N.N. Nikolaev[1,2,12], M. Nioradze[5], G. d'Orsaneo[1,2],
M. Papenbrock[10], D. Prasuhn[1,2], F. Rathmann[1,2], J. Sarkadi[1,2],
R. Schleichert[1,2], A. Smirnov[8], H. Seyfarth[1,2], J. Sowinski[11], D. Spoelgen[1,2],
G. Stancari[6], M. Stancari[3], M. Statera[3], E. Steffens[7], H.J. Stein[1,2],
H. Stockhorst[1,2], H. Straatmann[13,2], H. Ströher[1,2], M. Tabidze[5],
G. Tagliente[14], P. Thörngren Engblom[15], S. Trusov[16,17], A. Vasilyev[4],
Chr. Weidemann[1,2], D. Welsch[1,2], P. Wieder[1,2], P. Wüstner[9,2],
and P. Zupranski[18]

1. Institut für Kernphysik, Forschungszentrum Jülich GmbH, 52425 Jülich, Germany
2. Jülich Centre for Hadron Physics, 52425 Jülich, Germany
3. Università di Ferrara and INFN, 44100 Ferrara, Italy
4. St. Petersburg Nuclear Physics Institute, 188350 Gatchina, Russia
5. High Energy Physics Institute, Tbilisi State University, 0186 Tbilisi, Georgia
6. INFN, Sez. Ferrara, 44100 Ferrara, Italy
7. Physikalisches Institut II, Universität Erlangen–Nürnberg, 91058 Erlangen, Germany
8. Laboratory of Nuclear Problems, Joint Institute for Nuclear Research, 141980 Dubna, Russia
9. Zentralinstitut für Elektronik, Forschungszentrum Jülich GmbH, 52425 Jülich, Germany
10. Institut für Kernphysik, Universität Münster, 48149 Münster, Germany
11. Physics Department, Indiana University, Bloomington, IN 47405, USA
12. L.D. Landau Institute for Theoretical Physics, 142432 Chernogolovka, Russia
13. Zentralabteilung Technologie, Forschungszentrum Jülich GmbH, 52425 Jülich, Germany
14. INFN, Sez. Bari, 70126 Bari, Italy
15. Physics Department, Stockholm University, SE-106 91 Stocholm, Sweden
16. Institut für Kern- und Hadronenphysik, Forschungszentrum Rossendorf, 01314 Dresden, Germany
17. Skobeltsyn Institute of Nuclear Physics, Lomonosov Moscow State University, 119991Moscow, Russia
18. Department of Nuclear Reactions, Andrzej Soltan, Institute for Nuclear Studies, 00-681 Warsaw, Poland

*January 28, 2009*

**Abstract.** We discuss polarizing a proton beam in a storage ring, either by selective removal or by spin flip of the stored ions. Prompted by recent, conflicting calculations, we have carried out a measurement of the spin flip cross section in low-energy electron-proton scattering. The experiment uses the cooling electron beam at COSY as an electron target. The measured cross sections are too small for making spin flip a viable tool in polarizing a stored beam. This invalidates a recent proposal to use co-moving polarized positrons to polarize a stored antiproton beam.

## 1. INTRODUCTION

Usually, polarized ions in a storage ring are provided by injecting an already polarized beam from a suitable ion source. Alternatively, it is conceivable to polarize an initially unpolarized beam while it is stored in the ring. In the case of a spin-½ beam (with two spin states) this would be achieved by either selectively discarding particles in one spin state ("filtering"), or by selectively reversing the spin of particles in one spin state ("flipping").

After summarizing ideas and experimental results concerning the 'in-situ' polarization of a stored proton beam, we report in this paper a direct experimental evaluation of spin-flip in electron-proton scattering, and its contribution to polarizing the proton beam. The experiment, which is making use of the electron cooler as an electron target, has been carried out to resolve the discrepancy between two recently published calculations [1, 2], and to settle the question whether, in the future, spin flip will play a role in polarizing stored beams.

We are interested in the in-situ polarization of a stored proton beam because we hope to be able to apply the same technique to antiprotons. The need for polarized antiproton beams is well recognized as a prerequisite to address several important topics in particle physics, including a first direct measurement of the transversity distribution of the valence quarks in the proton, a test of the predicted opposite sign of the Sivers--function, related to the quark distribution inside a transversely polarized nucleon, and a first measurement of the moduli and the relative phase of the time-like electric and magnetic form factors of the proton [3].

Even though a number of methods to provide polarized antiproton beams have been proposed at a workshop more than 20 years ago [4], no polarized antiproton beams have yet been produced, with the exception of a low-quality, secondary beam from the decay of anti-hyperons that has been realized at Fermilab [5]. Recently, interest in the in-situ polarization of antiprotons has been revived, prompting workshops on the polarization of antiprotons at Daresbury [6], and at Bad Honnef [7].

## 2. IN-SITU POLARIZATION OF A STORED BEAM

### 2.1. Evolution of the beam polarization

Let us consider a storage ring that contains $N = N_\uparrow + N_\downarrow$ spin-½ particles in the two allowed substates, ↑ and ↓. The arrows indicate spins pointing along or opposite the quantization axis. The beam polarization is given by $P_B = (N_\uparrow - N_\downarrow)/N$. The beam interacts with an internal spin-½ target with polarization $P_T$ and area number density $d_T$. The orbit frequency is $f_R$.

A particle traversing the target may be removed from the stored beam by a reaction or by scattering at an angle larger than the ring acceptance. The removal cross section, integrated over the appropriate solid angle, is defined as $\sigma_R \equiv \frac{1}{2}(\sigma_{R(\downarrow\uparrow)} + \sigma_{R(\uparrow\uparrow)})$. The arrows indicate whether the spins of projectile and target are opposite or parallel. The spin-dependent part of the removal cross section is given by $\Delta\sigma_R \equiv \frac{1}{2}(\sigma_{R(\downarrow\uparrow)} - \sigma_{R(\uparrow\uparrow)})$.

In principle, it is also possible to change the polarization of the stored beam by spin flip of particles that interact with the target but remain in the ring. This mechanism would have the advantage that it entails *no beam loss*. The cross section for the spin flip of a beam particle is defined as $\sigma_S \equiv \frac{1}{2}(\sigma_{S(\downarrow\uparrow)} + \sigma_{S(\uparrow\uparrow)})$. The arrows indicate whether the spins of projectile and target (before the flip) are opposite or parallel. The spin-dependent part of the spin-flip cross section is given by $\Delta\sigma_S \equiv \frac{1}{2}(\sigma_{S(\downarrow\uparrow)} - \sigma_{S(\uparrow\uparrow)})$. These cross sections are integrated over the solid angle, from a minimum polar angle (given by Coulomb screening) to the angle that corresponds to the ring acceptance. Scattering within the ring acceptance, but *without* a spin flip, does not affect the beam polarization at all and can be ignored.

The time evolution equations for the beam polarization $P_B$ and the number of stored particles $N$ have been discussed repeatedly [8 - 12], and can be summarized as

$$\frac{dP_B}{dt} = f_R d_T \left[ 2P_T \Delta\sigma_S - 2\sigma_S P_B + P_T \left(1 - P_B^2\right)\Delta\sigma_R \right]. \tag{1}$$

$$\frac{dN}{dt} = -f_R d_T \left[ \sigma_R - P_T P_B \Delta\sigma_R \right] N. \tag{2}$$

It is useful to discuss two special cases. The first case deals with polarizing an initially unpolarized beam ($P_B = 0$). As long as $P_B$ is still small, the rate of change of polarization is constant and given by

$$\frac{dP_B}{dt} = f_R d_T P_T \left[ 2\Delta\sigma_S + \Delta\sigma_R \right]. \tag{3}$$

We define the 'polarizing cross section', $\sigma_{pol}$, as the sum of the two terms in the bracket.

The second special case describes the effect of an unpolarized target ($P_T = 0$) on an already polarized beam,

$$\frac{dP_B}{dt} = -2 f_R d_T \sigma_S P_B \ , \qquad (4)$$

which shows that the "de-polarizing cross section" is equivalent to twice the spin flip cross section $\sigma_S$. Since it is always true that $\sigma_S \geq \Delta\sigma_S$, it follows from eqs. 3 and 4 that *if a polarized target is capable of polarizing an unpolarized beam by spin flip, an unpolarized target will de-polarize an already polarized beam.* The experiment described in this paper makes use of this principle.

## 2.2. Spin filtering

The first (and so far only) evidence that a stored hadron beam can be polarized in situ was presented in 1993 by the FILTEX group [13]. The experiment was carried out in the TSR at Heidelberg with a 23-MeV proton beam, orbiting with $f_R = 1.177$ MHz in the presence of a polarized atomic hydrogen target. The target atoms were in a single spin state, i.e., protons and electrons were both polarized. The polarization buildup of an initially unpolarized beam was measured; the result was $dP_B/dt = (1.29 \pm 0.06)\cdot 10^{-2}$ per hour [13].

In the FILTEX experiment, the target thickness was $d_T = (5.3 \pm 0.3)\cdot 10^{13}$ cm$^{-2}$ and the target polarization was $P_T = 0.795 \pm 0.024$. Inserting these numbers into eq. 3, one finds for the polarizing cross section

$$\sigma_{pol} = (73 \pm 6) \text{ mb} \ . \qquad (5)$$

With positive target polarization the resulting beam polarization was positive, which means that the removal cross section is larger if beam and target spins are opposite, and thus $\sigma_{pol}$ is positive. The numbers of the final FILTEX result in eq. 5 slightly differ from those in the initial report [13]. These revisions followed the original publication within a year, and are discussed in [14].

When comparing the FILTEX result with the expectation based on the known pp interaction, it is important to realize that a beam ion has to scatter by at least an angle $\Theta_{acc}$ in order to leave the ring acceptance. For a well-cooled beam this angle is determined by the ring optics at the target, and, for the FILTEX experiment was measured as $\Theta_{acc} = (4.4 \pm 0.5)$ mrad [15]. For the transversely polarized target of the FILTEX experiment, the theoretical expectation for the polarizing cross section, assuming no spin flip effects (or, $\Delta\sigma_S = 0$), is [16]

$$\sigma_{pol,theor} = -2\pi \int_{\Theta_{acc}}^{\pi/2} \tfrac{1}{2}(A_{00nn} + A_{00mm}) \frac{d\sigma_0}{d\Omega} \sin\vartheta d\vartheta , \qquad (6)$$

where $A_{00nn}$ and $A_{00mm}$ are the spin correlation coefficients defined in ref. [17]. This expression has been evaluated numerically by a number of authors, using different parameterizations of the pp interaction [11,18,19]. Combining these results with a ±1 mb uncertainty, calculated from the uncertainty of the TSR acceptance angle, yields

$$\sigma_{pol,theor} = (86 \pm 2) \text{ mb} . \qquad (7)$$

The fact that experiment and theory (eqs. 5 and 7) disagree by two standard deviations has been the original motivation to investigate the role of spin flip.

### 2.3. Spin flip

During the analysis of the FILTEX result, it became clear that small-angle scattering, for which the ion remains in the beam, is a significant part of the total cross section [16]. It was argued that this scattering without loss may be accompanied by spin flip. This would include scattering not only from the polarized protons of the atomic beam target, but also from the electrons [20], which are also polarized. Because of their much larger mass, protons scattering from electrons always stay within the acceptance. Evaluating the spin *transfer* cross section at small angles between 10 and 100 MeV, sizeable effects were predicted [16]. The spin transfer cross section, as defined e.g. in ref. [17], refers to producing a spin-up beam particle (rather than a spin-down one) when an unpolarized beam interacts with a polarized target. A decade later, Milstein and co-workers [10] showed that the relevant quantity to evaluate is the spin *flip* cross section (the cross section that the spin of a beam particle is reversed.), which is different from, and much smaller that the spin transfer cross section and is in fact negligible for the proton energy used in the FILTEX experiment.

More recently, Arenhövel [1] predicted that the spin-flip cross section in electron-proton scattering at low energy (a few eV in the center-of-mass system) is very large because of the mutual attraction of the two oppositely charged particles. Walcher and co-workers adopted this idea for a proposal to polarize stored antiprotons with a co-moving beam of polarized positrons [21]. The proper low interaction energy would be achieved by making the two beam velocities almost the same. Even though the achievable positron beam intensities are quite low, the predicted spin flip cross sections are so large that the scheme would still be feasible. For instance, at a center-of-mass energy of 0.93 eV (corresponding to a proton energy in the lepton rest frame of $T_h = 1.7$ keV) Arenhövel predicts a spin flip cross section of $\sigma_S = 4 \cdot 10^{13}$ b. However, a calculation of the same quantity by Milstein and co-workers [2] resulted in $\sigma_S = 0.75$ mb. The goal of the experiment described in the following is to resolve this discrepancy of 16 orders of magnitude.

## 3. EXPERIMENT

The aim of this experiment is to determine the depolarization of a polarized proton beam by its interaction with the electrons of the cooler beam. The measurement is carried out with a proton beam in the COSY ring [22], using the detector setup in the target chamber of the ANKE spectrometer [23]. The proton energy is $T_p = (49.3 \pm 0.1)$ MeV, corresponding to a velocity of $v_p = 0.312 \cdot c$, where $c$ is the speed of light and $\gamma_p = 1.053$ is the usual relativistic parameter.

### 3.1. Cooler beam as an electron target

In this experiment the COSY electron cooler [24] serves two functions. On one hand, as usual, it provides the phase-space cooling of the stored proton beam, while on the other hand it plays the role of an electron target for the actual measurement of the low-energy spin-flip cross section in e-p scattering.

In the cooling mode, the electron velocity $v_e$ is adjusted to the velocity $v_p$ of the stored protons. This is the case if the accelerating potential equals $U_C = (m_e / m_p) T_p$, where $m_e$ and $m_p$ are the particle masses, and $T_p$ is the proton kinetic energy. When the cooler is used as a target, a relative motion between the proton and the electron beam is achieved by 'detuning' the accelerating voltage by $\Delta U$, changing the electron velocity by $\Delta v_e$, and inducing an average relative 'detune' velocity $u_0$.

Besides this induced velocity, there are additional contributions to the relative motion between protons and electrons. The dominant effect arises from the transverse thermal motion of the electrons. Other contributions include the betatron motion of the protons, the velocity spread of both beams, and the ripple on the electron high-voltage supply.

#### 3.1.1. Average over the relative electron-proton velocities

In the following, we only deal with the transverse motion of the electrons, having ascertained that the remaining effects are at least a factor of 10 smaller. We thus assume that the distribution of relative electron-proton velocities, $\vec{u}$, taking into account all possible pairings, is the combination of the detune velocity $u_0$ with a Maxwell distribution in the transverse direction, given by

$$g(\vec{u}) d^3\vec{u} = \frac{1}{2\pi\xi^2} e^{-\frac{u_r^2}{2\xi^2}} \delta(u_z - u_0) u_r \, du_r \, d\varphi \, du_z \ . \tag{8}$$

Here, cylindrical coordinates $(r, z, \varphi)$ are used. The distribution $g(\vec{u})$ is normalized to 1. We choose a coordinate system, the 'proton rest frame', which is moving with $v_p$, with the z-axis in the beam direction, the y-axis up, and the x-axis in the ring plane, i.e., $x = r$

cos$\varphi$ and $y = r$ sin$\varphi$. The parameter $\xi = \gamma_p \sqrt{kT_e/m_e}$ is determined by the transverse electron temperature $T_e$. Velocities are given in units of $c$, and the detune velocity is $u_0 = \gamma_p^2 \Delta v_e$.

The expression, eq. 4, for the beam depolarization then becomes

$$\frac{dP}{dt} = -2\frac{L_C}{L_R} n_e \langle u\sigma_S(u) \rangle \cdot P \ . \tag{9}$$

From now on, $P$ signifies the beam polarization, $L_c$ is the active length of the cooler, $L_R$ the ring circumference, $n_e$ the electron number density, and the angular bracket represents an average over the velocity distribution $g(\vec{u})$. In principle, eq. 9 is evaluated in the proton rest frame, however the product $n_e \cdot dt$ is Lorentz-invariant, allowing us to use laboratory values for time and electron density. The latter is given by

$$n_e = I_e /(e A_e v_e c) \ , \tag{10}$$

where $e$ is the elementary charge, $v_e$ the lab velocity of the electrons, $I_e$ the electron current, and $A_e$ the cross-sectional area of the electron beam.

The task at hand is to evaluate the average over the distribution of relative velocities. An additional complication arises from the fact that there are two independent depolarizing cross sections, $\sigma_{S,\tau}$ and $\sigma_{S,\lambda}$, depending on whether the polarization vector is transverse to or along the motion $\vec{u}$. In this experiment the polarization is along the y-axis, and thus the average over $\vec{u}$ is given by

$$\langle u\sigma_S(u) \rangle = \int u \cdot \left( \sigma_{S,\tau}(u) \frac{u_r^2 \cos^2\varphi + u_z^2}{u^2} + \sigma_{S,\lambda}(u) \frac{u_r^2 \sin^2\varphi}{u^2} \right) \cdot g(\vec{u}) \, d^3\vec{u} \ . \tag{11}$$

Inserting $g(\vec{u})$, integrating over $u_z$ and $\varphi$, and replacing $u_r$, using $u_r^2 = u^2 - u_0^2$ and $u_r \, du_r = u \, du$, leads to

$$\langle u\sigma_S(u) \rangle = \frac{1}{2\xi^2} \int_{u_0}^1 \left[ \sigma_{S,\tau}(u)\left(1 + \frac{u_0^2}{u^2}\right) + \sigma_{S,\lambda}(u)\left(1 - \frac{u_0^2}{u^2}\right) \right] \cdot e^{-\frac{u^2 - u_0^2}{2\xi^2}} u^2 \, du \ . \tag{12}$$

For the small relative velocities considered here, the spin-flip cross sections are to a good approximation proportional to $u^{-2}$. This velocity dependence is discussed by Milstein et al. [10], and follows from their eq. 20, when neglecting the logarithmic term (which amounts to just a few percent). We thus set

$$\sigma_{S,j}(u) = \frac{u^{*2}}{u^2} \sigma^*_{S,j} \quad , \tag{13}$$

where $j$ stands for either $\tau$ or $\lambda$, and $u^*$ is an arbitrarily chosen reference velocity with the understanding that the experiment yields information on the two cross sections $\sigma^*_{S,\tau}$ and $\sigma^*_{S,\lambda}$, *at that velocity*. The velocity average can now be written as

$$\langle u \sigma_S(u) \rangle = u^{*2} \left( \sigma^*_{S,\tau} I_\tau + \sigma^*_{S,\lambda} I_\lambda \right) \quad , \tag{14}$$

with

$$I_j(u_0) = \int_{u_0}^{1} w_j(u, u_0) \, du \quad , \tag{15}$$

and

$$w_j(u, u_0) = \frac{1}{2\xi^2} \left( 1 \pm \frac{u_0^2}{u^2} \right) e^{-\frac{u^2 - u_0^2}{2\xi^2}} \quad . \tag{16}$$

The weight functions $w_j$ are defined for $u > u_0$; the plus sign applies when $j = \tau$, and the minus sign when $j = \lambda$. The integrals $I_\tau$ and $I_\lambda$, evaluated numerically for a transverse electron temperature $kT_e = 0.3$ eV ($\xi = 8.1 \cdot 10^{-4}$), are listed in tab. 1. Although not needed for the subsequent data analysis, it is interesting to calculate an average velocity $\bar{u}$ and range $\Delta\bar{u}$ for the relative velocity from the centroid and width of the weight functions (the corresponding values are very similar for the two weight functions, so we only quote their mean).

| $\Delta U$ | $u_0$ | $I_\tau$ | $I_\lambda$ | $\bar{u}$ | $\Delta\bar{u}$ | $R$ | $\delta R$ |
|---|---|---|---|---|---|---|---|
| (V) | ($10^{-3} c$) | | | ($10^{-3} c$) | ($10^{-3} c$) | | |
| −426 | −2.53 | 338 | 25.8 | 2.82 | 0.23 | 1.084 | 0.051 |
| 0 | 0 | 777 | 777 | 0.64 | 0.49 | 1.022 | 0.055 |
| 246 | 1.46 | 482 | 79.4 | 1.86 | 0.30 | 1.008 | 0.096 |
| 301 | 1.79 | 429 | 54.3 | 2.15 | 0.27 | 1.094 | 0.074 |
| 348 | 2.07 | 391 | 40.3 | 2.40 | 0.25 | 1.023 | 0.055 |
| 426 | 2.53 | 338 | 25.8 | 2.82 | 0.23 | 1.027 | 0.049 |

**Table 1.** The detune voltages $\Delta U$ used in this experiment, the detune velocity $u_0$ in the proton rest frame, the integrals $I_\tau$ and $I_\lambda$ of eq. 15, and the average velocity $\bar{u}$ and its standard deviation. The calculation assumes a transverse electron temperature of $kT_e = 0.3$ eV. The last two columns show the measured ratio $R$, of the polarization with and without electron target and its statistical uncertainty $\delta R$ (see sect. 3.3.3).

### 3.1.2. *Effect of the detuned electrons on the stored beam*

Whenever the detune potential is set to a non-zero value, the stored beam is no longer cooled. In addition, since no RF bunching is used, the drag force exerted by the electrons will affect the proton beam velocity. In order to limit this effect, we alternate 5 s intervals of detuned operation ($\Delta U \neq 0$) with 5 s intervals of cooling (see sect. 3.2).

In order to demonstrate that the drag force does not significantly affect the proton velocity during the detuned phase (5 s) we show in fig. 1 the frequency distribution of the orbiting particles (Schottky spectrum). The top trace shows the spectrum of the *cooled* beam with the characteristic double peak due to plasma waves traveling in opposite directions. The remaining four traces are measured after a detune voltage of $\Delta U = 246$ V has been turned on, which results in a relative electron-proton velocity of $u_0 = 1.46 \cdot 10^{-3}$. The bottom scale shows the proton velocity $u_p$ relative to its value at the start of the detune interval. One notices that after applying the detune voltage, the proton velocity spread immediately increases, and continues to grow slightly while the centroid shifts by about $0.08 \cdot 10^{-3}$ over the 10 s interval studied here. In the actual experiment, the detune interval is 5 s. It is clear that both spread and shift of the proton velocity distribution are negligible compared to $u_0$.

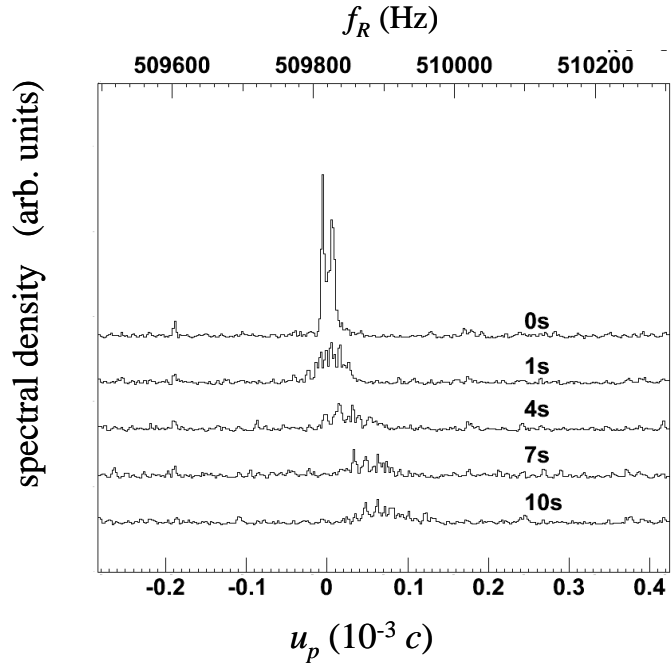

**Fig. 1:** Schottky spectra (distribution in orbit frequency) of the stored proton beam. The top trace is measured when the beam is cooled. The remaining traces show the evolution of the distribution after a detune voltage of 246 V has been applied. The bottom scale shows the proton velocity $u_p$ relative to its value at the start of the detune interval. The measurement covers a period of 10 s.

## 3.2. Cycle scenario

An experiment with a storage ring typically consists of a sequence of fill cycles. The scenario of our experimental cycle is shown in fig. 2. At the beginning of the cycle, the ring is filled with vertically polarized protons (typically, the beam polarization is $P_B \sim 0.5$). During the first half of the cycle, the coasting beam is interacting with the electrons in the cooler. During the second half, while cooling the beam, the internal deuteron target is turned on to measure the beam polarization.

The first half of the cycle contains 49 sub-cycles of 10 s length. During such a sub-cycle the electron velocity is first tuned to the beam velocity to cool the beam for 5 s, then the electron beam velocity is detuned for another 5 s. This is the time when the actual experiment takes place with a total 'interaction' time in the detuned mode of $t_{int} = 245$ s per cycle.

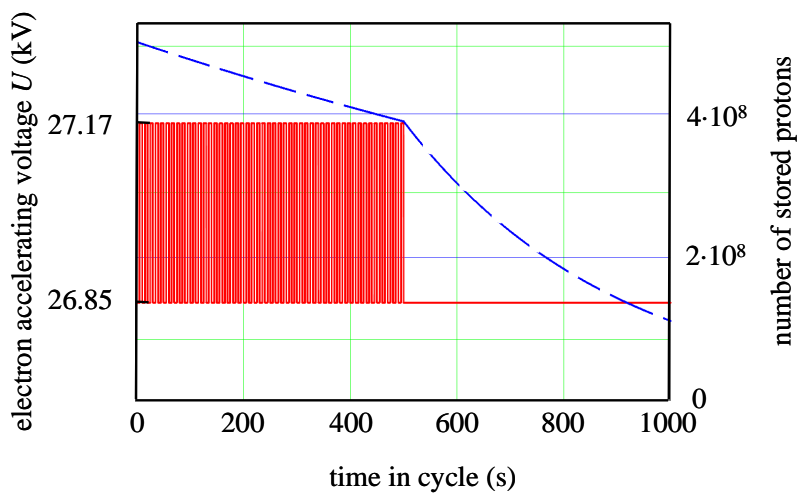

**Fig. 2:** Time sequence of an experimental cycle. The lower trace (solid line) shows the electron accelerating voltage, in the first half alternating between cooled and detuned mode. In this case, the cooling electron accelerating voltage is $U_C = 26.8$ kV, and the detune voltage is $\Delta U = 320.42$ V. The upper trace (dashed) indicates the number of protons orbiting in the ring. The deuterium target is turned on after 500 s.

The scenario just described shall be called '*E*-cycle'. To reduce systematic uncertainties, *E*-cycle polarization measurements are compared to those observed in a reference cycle, or '*0*-cycle'. Reference cycles are identical in every respect, except that during the interaction time (in the second half of the sub-cycles, for a total time $t_{int}$ in each cycle) the cooler beam is *turned off*. During the experiment, *E*-cycles and *0*-cycles are alternated, first with beam polarization up (↑), then with an unpolarized beam and finally with polarization down (↓). This sequence is repeated 30 to 40 times to acquire statistics. The deduced polarization ratio $R \equiv P_E/P_0$ (see sect. 3.3.3) reflects the effect of an electron target on the beam polarization. The data obtained with an unpolarized beam are used for systematic studies.

### 3.3. Polarimetry

#### 3.3.1. Hardware

The beam polarization is measured using p+d elastic scattering. Precise analyzing power data are available at $T_p = 49.3$ MeV [25] and cross sections have been measured at a nearby energy ($T_p = 46.3$ MeV) [26]. The beam energy for this experiment was chosen partly because of this.

The target consists of a deuterium cluster jet with about $5 \cdot 10^{14}$ deuterons per cm$^2$ [27]. The target thickness has been measured, using the beam energy loss, which in turn is deduced from a shift of the orbit frequency of the coasting beam [28].

The detector system consists of two silicon tracking telescopes [29] placed symmetrically to the left and right of the beam. Each telescope features three position-sensitive detectors, oriented parallel to the beam direction. The first two layers are 300 µm thick with an active area of 51 mm by 66 mm. They are located 30 mm and 50 mm from the beam axis. The third, 5 mm thick detector, 70 mm from the beam axis is not used in this experiment. Within the mechanical constraints of the detector support, the telescope positions with respect to the interaction region are chosen to optimize the figure of merit for the p+d analyzing reaction. The position resolution of the detectors is about 200 µm, both, vertically (y axis) and along the beam direction (z axis).

#### 3.3.2. Event selection

A scatter plot of the energy deposited in the first two layers of one of the telescopes is shown in fig. 3. Deuterons, most of them stopping in the second layer, can be selected cleanly by a single cut (solid line in fig.3). The remaining part of the deuteron locus (below the solid line) overlaps with the proton locus. Deuterons in this part of the locus are also identified, provided that they are accompanied by a coincident proton in the other telescope, and that the two tracks are co-planar with the beam axis, and the respective polar angles are consistent with the kinematics of the p+d reaction. Events recovered in this manner amount to 11% of the number of events above the cut. The polarizations deduced separately from events above and below the cut are consistent within statistics. The total number of processed events for each of the six detune voltages is between 230'000 and 760'000. The efficiency of the polarimeter, i.e., the fraction of scattering events that lead to an identified deuteron, is about 1 %.

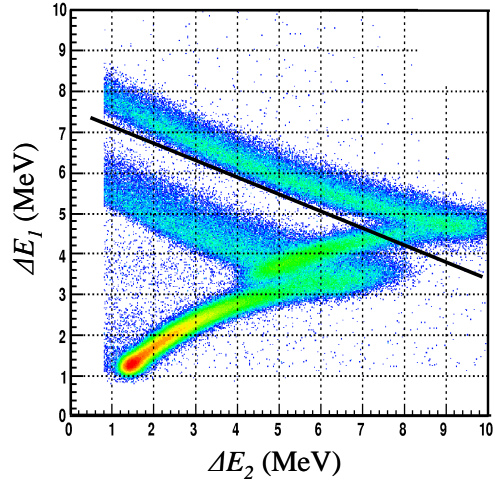

**Fig. 3:** Deposited energy in the second layer versus that in the first layer. The solid line indicates the primary cut to identify deuterons to be processed. Some deuterons below the line are identified as well and included in the analysis (see text).

Events, in which both, the proton and the deuteron, have been observed, are also used to reconstruct the reaction vertex by combining the track information with the known kinematics of the event. This is used to demonstrate that the beam position is stable to within ±10 μm throughout the experiment, excluding beam shifts as a possible source of false asymmetries.

### 3.3.3. Determination of the beam polarization

The selected events are sorted into bins covering $\theta_n \pm 1.5°$, where $\theta$ is the laboratory deuteron scattering angle and $n$ is the bin number. This is done separately for the left and the right detector and for runs with up or down beam polarization (↑,↓), resulting in the four yields $Y_{L\uparrow}(n)$, $Y_{R\downarrow}(n)$, $Y_{R\uparrow}(n)$, $Y_{L\downarrow}(n)$. Making use of the cross ratio method [30], we calculate the asymmetry for each angle bin

$$\varepsilon_n = \frac{1}{\langle \cos\varphi \rangle} \frac{\delta_n - 1}{\delta_n + 1} \quad , \tag{17}$$

where

$$\delta_n = \sqrt{\frac{Y_{L\uparrow}(n) \cdot Y_{R\downarrow}(n)}{Y_{L\downarrow}(n) \cdot Y_{R\uparrow}(n)}} \quad . \tag{18}$$

The average $\langle \cos\varphi \rangle$ over the azimuthal coverage of the detector takes into account the dependence of the analyzing power on azimuth. Asymmetries obtained during a typical run are shown in fig.4. The solid curve results from a polynomial fit to the known analyzing power [25], folded with the width of the angle bins, and scaled to fit the data in the figure.

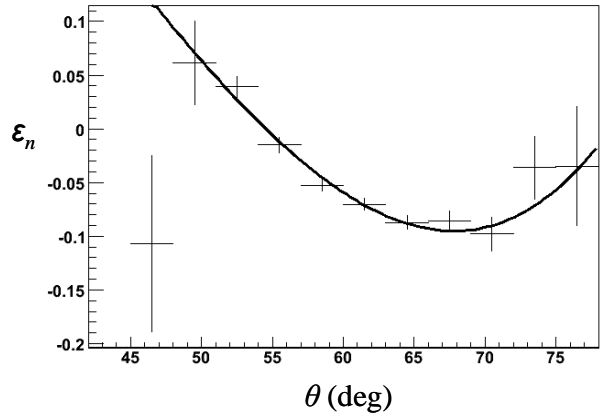

**Fig. 4**: Asymmetries $\varepsilon_n$ with their statistical uncertainty versus the deuteron laboratory angle observed during the run with $\Delta U = 246$ V. The horizontal bars indicate the bin width (3°). The curve is deduced from the known analyzing power (see text).

For each angle bin the analyzing power $\overline{A}_{y,n}$ that represents the data from [25] is calculated, using the polynomial fit and the measured $\theta$ for all events in that bin. Each bin then yields the value $\varepsilon_n / \overline{A}_{y,n}$ for the beam polarization. Taking the weighted average for all bins, one arrives at the overall beam polarization. This procedure is carried out separately for *E*-cycles and *0*-cycles, resulting in the respective polarizations $P_E$ and $P_0$, with or without electron beam during the 'interaction' part of the cycle. The ratio $R \equiv P_E/P_0$ then constitutes the final result of the polarization measurement. The values for $R$ and the statistical uncertainties $\delta R$ for the six detune velocities are listed in tab. 1, and plotted in fig. 5. It is believed that the systematic errors of this measurement can be neglected, since the beam position was stable, the up and down polarizations were the same within statistics, and systematic asymmetries in beam current and target density cancel to first order in the cross ratio. Furthermore, the ratio $R$ depends only on the *change* of the polarization between *E*-cycles and *0*-cycles, and the actual value of the beam polarization (between 0.47 and 0.53) merely affects $\delta R$, while the normalization of the imported analyzing power cancels.

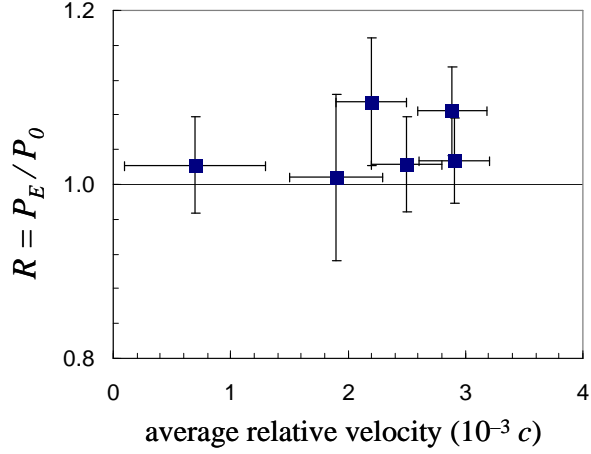

**Fig. 5:** Ratio of the beam polarization with or without electron beam during the 'interaction' part of the cycle as a function of the average relative velocity $\bar{u}$. The horizontal bars indicate the range $\Delta\bar{u}$ of velocities that contribute to the measurement. The vertical bars are statistical uncertainties.

## 4. RESULT

For each of the six detune potentials $\Delta U_k$ ($k = 1\ldots6$) (see tab. 1), the result of the measurement consists of the ratios $R_k \equiv (P_E/P_0)_k$ as described in the previous section. When combining eqs. 9 and 14 one obtains

$$y_k \equiv \frac{-\ln R_k}{2\, c\, t_{int}\, n_{e,k}\, u^{*2}\, (L_C/L_R)} = \sigma^*_{S,\tau}\, I_{\tau,k} + \sigma^*_{S,\lambda}\, I_{\lambda,k} \quad . \tag{19}$$

The denominator contains the speed of light, the interaction time $t_{int} = 245$ s, the electron density $n_e$, the reference velocity (see eq. 13), arbitrarily set to $u^* = 0.002$, the active length $L_C = (1.75 \pm 0.25)$ m of the cooler, and the ring circumference $L_R = 183.47$ m. The cooler length is uncertain because of details of inflection and extraction of the electron beam, and the electron density (eq. 10) is affected by uncertainties of the electron beam current $I_e = 170$ mA and its area $A_e = 5$ cm$^2$. We estimate that the overall systematic uncertainty of the denominator is $\pm 20$ %.

The polarization ratios $R_k$ (fig.5) are consistent with unity, i.e., the polarization differences between *E*-cycles and *0*-cycles are of the order of their statistical errors. A possible systematic effect arises from the fact that the negative charge distribution of the cooling electrons slightly focuses the proton beam. This affects the machine tune, and thus the effect of nearby depolarizing resonances. Consequently, the polarization lifetime during the interaction time in *E*-cycles and *0*-cycles could differ slightly, because of the absence of electrons in the latter, which would affect the measured ratio $R_k$. An indication of this effect might be seen in the fact that all $R_k$ are larger than unity, corresponding to a larger polarization *with* the electron beam than without it. Such an effect is opposite from that of a possible beam depolarization, and it is conceivable that both are sizeable, but compensate each other. However, such compensation could only be achieved for a single measured detune velocity, while all other ratios $R_k$ would deviate significantly from unity, which is not the case. We estimate this systematic uncertainty of $R_k$ to be $\pm 0.05$.

The integral weights $I_{\tau,k}$ and $I_{\lambda,k}$ on the right side of eq. 19 depend on the transverse electron temperature $kT_e$, which, for a number of reasons, is larger than the temperature of the emitting cathode, which operates at 900 °C, corresponding to $kT_e = 0.1$ eV. The actual temperature, $kT_e = (0.3 \pm 0.1)$ eV, has been deduced from a measurement of the rate of electron pick-up by co-moving protons [31].

The depolarizing cross sections, $\sigma^*_{S,\tau}$ and $\sigma^*_{S,\lambda}$ (at the reference velocity $u^*$) appear as unknowns in eq. 19. Since our experiment fails to find a depolarization effect, we instead derive an upper limit for the two cross sections that is compatible with our data. Following the usual treatment (see, e.g., [32]), we define the likelihood function

$$L(\vec{y}|\sigma^*_{S,\tau},\sigma^*_{S,\lambda}) \equiv \prod_k \exp\left(\frac{(y_k - \sigma^*_{S,\tau} I_{\tau,k} - \sigma^*_{S,\lambda} I_{\lambda,k})^2}{2\delta y_k^2}\right), \quad (20)$$

The experimental result, $y_k$, is defined in eq. 19; the statistical uncertainty $\delta y_k$ follows from the error $\delta R$ (see tab. 1). Following the Bayesian approach, we calculate the posterior probability density function

$$p(\sigma_{S,\tau},\sigma_{S,\lambda}|\vec{y}) = \frac{L(\vec{y}|\sigma^*_{S,\tau},\sigma^*_{S,\lambda}) h(\sigma^*_{S,\tau},\sigma^*_{S,\lambda})}{\int L(\vec{y}|\hat{\sigma}^*_{S,\tau},\hat{\sigma}^*_{S,\lambda}) h(\hat{\sigma}^*_{S,\tau},\hat{\sigma}^*_{S,\lambda}) d\hat{\sigma}^*_{S,\tau} d\hat{\sigma}^*_{S,\lambda}}, \quad (21)$$

The function $h$ reflects our prior knowledge (cross sections are positive numbers) and is set to a constant for all non-negative values of $\sigma^*_{S,\tau}$ and $\sigma^*_{S,\lambda}$, and to zero otherwise.

The probability $p$ is evaluated numerically. The upper cross section limits, shown in fig.6, are contours of constant $p$. The significance level (S.L.) is the integral of $p$ over the region below the curve, and equals the probability that the two cross sections are less than the values along the contour. When evaluating the integral, we assume that parameters with a systematic uncertainty are completely unknown within a range equal to that uncertainty. We thus conservatively choose a value for these parameters (within that range) that result in the largest upper limit.

As mentioned earlier, the spin flip cross sections are proportional to the inverse square of the relative velocity $u^*$. The values shown in fig.6 are for $u^* = 0.002$, corresponding to a center-of mass energy of about 1 eV, or to a proton kinetic energy in the electron rest system of $T_h = 1.2$ keV.

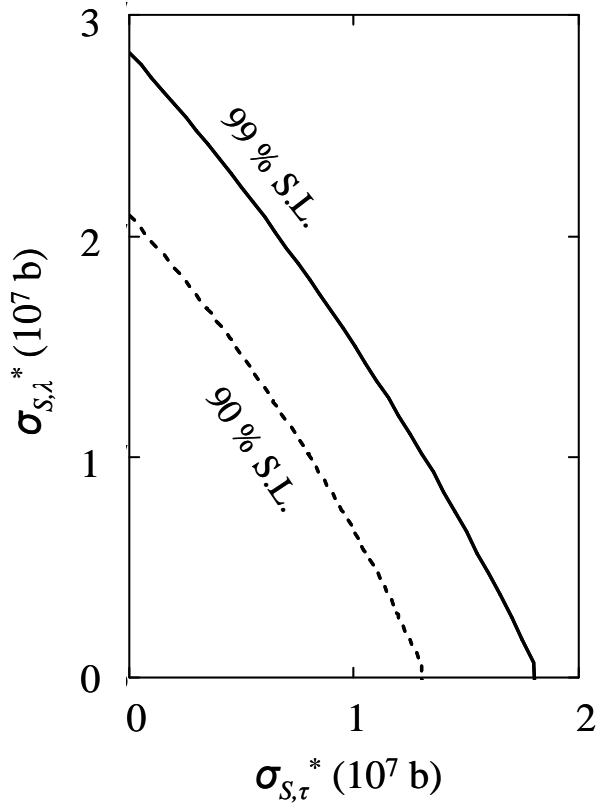

**Fig. 6:** Upper limit allowed by the data of this experiment for the transverse and the longitudinal spin flip cross sections $\sigma^*_{S,\tau}$ and $\sigma^*_{S,\lambda}$ at a relative velocity of $u^* = 0.002$, corresponding to a center-of mass energy of about 1 eV. The significance level (S.L.) is the probability that the actual cross sections are smaller than the values on the contour line.

## 5. CONCLUSIONS

We have reviewed the history, motivation and previous attempts to polarize a stored beam in situ. This includes a discussion of the formalism for polarization evolution, taking into account both, filtering and spin flip.

In a dedicated experiment we have searched for the depolarization of a stored proton beam due to spin-flip in electron-proton scattering at relative velocities between $6.4 \cdot 10^{-4}$ $c$ and $2.8 \cdot 10^{-3}$ $c$, which corresponds to center-of-mass energies between 100 meV and 2 eV. The low interaction energies are accomplished by using overlapping proton and electron beams, moving together at slightly different velocity.

The experiment is sensitive to spin flip with either the polarization vector along the relative motion, or transverse to it. Both corresponding spin flip cross sections, $\sigma_{S,\lambda}$ and $\sigma_{S,\tau}$ are found to be consistent with zero with an upper limit (at $u^* = 0.002$, or 1 eV c.m. energy) of the order of $10^7$ b (for details, see fig. 6). Since the measured spin-flip cross section $\sigma_S$ is always larger than the spin-dependent $\Delta\sigma_S$, the present experiment rules out the usefulness of spin flip in low-energy electron-proton or positron-antiproton scattering in polarizing a stored beam.

The present result is in agreement with the calculation of Milstein et al. [2], but clearly rules out the validity of the prediction of $\sigma_{S,\lambda} = 4\cdot10^{13}$ b claimed in refs. [1,21]. Since the completion of this experiment, the calculation presented in these two references has been withdrawn [33,34]).

Thus, it now seems that in-situ polarization of antiprotons will have to rely exclusively on the spin filtering mechanism. Theoretical estimates of the filtering effect of a polarized hydrogen target in a stored antiproton beam predict a significant polarization build-up for beam energies between 20 and 100 MeV [35], calling for an experimental effort to verify this prediction.

The PAX collaboration [36] is currently pursuing a program to this effect [35]. The proposed program includes a measurement of spin-correlation observables in proton-antiproton scattering, using the antiproton beam of the AD ring at CERN at antiproton beam energies in the range from 50 to 200 MeV in conjunction with an internal polarized hydrogen storage cell gas target. The data will allow the definition of the optimum working parameters of a dedicated Antiproton Polarizer Ring (APR), which has recently been proposed by the PAX collaboration for the new Facility for Antiproton and Ion Research (FAIR) at GSI in Darmstadt, Germany.

## 6. ACKNOWLEDGMENTS

We are grateful to the operators of the COSY facility for their help in setting up the accelerator and the unusual performance parameters of the cooler. We also appreciate the contributions of a number of members of the ANKE and PAX collaborations who were not directly involved in the experiment. The present work was supported by grants of the JCHP-FFE program, HGF-VIQCD, BMBF Verbundforschung, the BMBF-Dubna grant, the DFG, INFN, NSF, and the Georgian National Science Foundation. H.O. Meyer would like to thank the Humboldt Foundation for their support which made his participation in this experiment possible.

# 7. REFERENCES

[1] H. Arenhövel, Eur. Phys. J. A 34 , 303 (2007).
[2] A.I. Milstein, S.G. Salnikov and V.M. Strakhovenko, Nucl. Instr. Meth. B266, 3453 (2008).
[3] Technical Proposal for Antiproton-Proton Scattering Experiments with Polarization, PAX Collaboration, http://arxiv.org/abs/hep-ex/0505054 (2005). An update can be found at the PAX website http://www.fz-juelich.de/ikp/pax
[4] Proc. Workshop on Polarized Antiprotons, Bodega Bay, CA, 1985, (A.D. Krisch, A.M.T. Lin and O. Chamberlain, eds), AIP Conf. Proc. 145 (1986).
[5] D.P. Grosnick et al., Nucl. Instr. Meth. A 290, 269 (1990).
[6] Proc. Int. Workshop on Polarized Antiproton Beams, (D.P. Barber, N. Buttimore, S. Chattopadhyay, G. Court and E. Steffens, eds.) AIP Conf. Proc. 2008.
[7] 409th WE-Heraeus Seminar on Polarized Antiprotons, Bad Honnef, 23-25 June 2008, http://www.fe.infn.it/heraeus/index.html.
[8] N.H. Buttimore and D.S. O'Brien, Eur. Phys. J. A35, 47 (2008).
[9] W.W. MacKay and C. Montag, Phys. Rev. E 73, 028501 (2006).
[10] A.I. Milstein and V.M Strakhovenko, Phys. Rev. E 72, 066503 (2005).
[11] N. Nikolaev and F. Pavlov, http://arXiv.org/abs/hep-ph/0701175.
[12] D.S. O'Brien, http://arXiv.org/abs/0711.4819.
[13] F. Rathmann et al., Phys. Rev. Lett. 71,1379 (1993).
[14] F. Rathmann, Ph. D. thesis, Phillips-Universität Marburg, Jan. 1994.
[15] K. Zapfe et al., Nucl. Instr. Meth. A368, 293 (1996).
[16] H.O. Meyer, Phys. Rev. E50, 1485 (1994).
[17] J. Bystricky, F. Lehar and P. Winternitz, J. Phys. (Paris) 39, 1 (1978) .
[18] V. Strakhovenko, in Proc. Int. Workshop on Polarized Antiproton Beams, ref. [DAR08], p.44
[19] N. Nikolaev and F. Pavlov, in Proc. Int. Workshop on Polarized Antiproton Beams, ref. [DAR08], p.34
[20] C. Horowitz and H.O. Meyer, Phys. Rev. Lett. 72, 3981 (1994).
[21] Th. Walcher et al., Eur. Phys. J. A 34 , 447 (2007).
[22] R. Maier et al., Nucl. Instr. Meth. A390, 1 (1997).
[23] S. Barsov et al., Nucl. Instr. Meth. A462, 364 (2001).
[24] H.J. Stein et al., Atomic Energy 94, 24 (2003) and Proc. XVIII Conference on Accelerators of Charged Particles, RUPAC-2002, Obninsk, Russia, (I.N. Meshkov et al., eds.) Obninsk 2004, Vol. 1, p. 220.
[25] N.S.King et al., Phys. Lett. B 69, 151 (1977).
[26] S.N. Bunker et al., Nucl. Phys. A 113, 461 (1968).
[27] A. Khoukaz et al., Eur. Phys. J. D 5, 275 (1999).
[28] H.J. Stein et al., Phys. Rev. ST-AB, 11, 052801 (2008).
[29] R. Schleichert et al., IEEE Trans. Nucl. Sci. 50, 301 (2003).
[30] R.C. Hanna, Proc. 2nd Int. Symp. on Polarization Phenomena, Karls ruhe 1965, Birkhaeuser Basel 1966, p. 280.
[31] H. Poth, Physics Reports 196, 135 (1990).
[32] C. Amsler et al., Phys. Lett. B667, 1 (2008).
[33] H. Arenhövel, Eur. Phys. J. A 39 , 133 (2009).
[34] Th. Walcher et al., Eur. Phys. J. A 39 , 137 (2009).
[35] V.F. Dmitriev, A.I. Milstein and V.M. Strakhovenko, Nucl. Instr. Meth. B 266, 1122 (2008).
[36] PAX collaboration, spokespersons: P. Lenisa (Ferrara University, Italy) and F. Rathmann (Forschungszentrum Juelich, Germany), http://www.fz-juelich.de/ikp/pax.
[37] Letter-of-Intent for Measurement of the Spin-Dependence of the pp Interaction at the AD-Ring, PAX Collaboration, http://arxiv.org/abs/nucl-ex/0512021 (2005).